\documentclass[english,aps,prl,twocolumn, 16pt, longbibliography]{revtex4-1}
\usepackage[dvips]{graphicx}
\usepackage{dcolumn}
\usepackage{bm}
\usepackage{float}
\usepackage{hyperref}
\usepackage{color}
\usepackage{cancel}
\usepackage{bmpsize}
\usepackage{amsmath}
\usepackage{amssymb}
\usepackage{natbib}
\usepackage{wrapfig}

\newcommand{\bra}[1]{\left\langle #1\right|}
\newcommand{\ket}[1]{\left| #1\right\rangle}

\newcommand{\ketbra}[2]{\left| #1\right\rangle\!\left\langle#2\right|}
	

\begin{document}

\title{Entanglement of macroscopically distinct states of light}

\author{Demid V. Sychev$^{1,4}$, Valeriy A. Novikov$^{1,2}$,  Khurram K. Pirov$^{1,2}$, Christoph Simon$^{5}$, and A.I. Lvovsky$^{1,3,5,6}$\vspace{5mm}}

 \affiliation{$^1$Russian Quantum Center, 100 Novaya St., Skolkovo, Moscow 143025}
 \affiliation{ $^2$Moscow Institute of Physics and Technology, 141700 Dolgoprudny}
\affiliation{$^3$P. N. Lebedev Physics Institute, Leninskiy prospect 53, Moscow 119991, Russia}
\affiliation{$^4$Moscow State Pedagogical University, Department of Theoretical Physics, M. Pirogovskaya Street 29, Moscow 119991, Russia}
\affiliation{$^5$Institute for Quantum Science and Technology, University of Calgary, Calgary AB T2N 1N4, Canada}
\affiliation{$^6$Department of Physics, University of Oxford, Oxford OX1 3PU, UK}
\email{Alex.Lvovsky@physics.ox.ac.uk}

\date{\today}

\begin{abstract}
Schr\"odinger's famous \emph{Gedankenexperiment} has inspired multiple generations of physicists to think about apparent paradoxes that arise when the logic of quantum physics is applied to macroscopic objects. The development of quantum technologies enabled us to produce physical analogues of Schr\"odinger's cats, such as superpositions of macroscopically distinct states as well as entangled states of microscopic and macroscopic entities. Here we take one step further and prepare an optical state which, in Schr\"odinger's language, is equivalent to a superposition of two cats, one of which is dead and the other alive, but it is not known in which state each individual cat is. Specifically, the alive and dead states are, respectively, the displaced single photon and displaced vacuum (coherent state), with the magnitude of displacement being on a scale of $10^8$ photons. These two states have significantly different photon statistics and are therefore macroscopically distinguishable.
\end{abstract}

\maketitle

\paragraph{Introduction.} Counterintuitive quantum effects such as superposition and entanglement can usually only be observed at the microscopic scale, but it is interesting for a variety of reasons to try to bring them to the macroscopic level \cite{frowis}. First, the applicability range of quantum theory is one of the big unresolved questions of modern physics. If quantum physics indeed applies to macroscopic systems, including ourselves, then this implies the existence of parallel universes. The alternative is that quantum principles cease to hold at some level of macroscopicity, in which case it is important to probe the underlying physics \cite{arndt-hornberger,marshall}. Second, bringing quantum phenomena to macroscopic scales would make them directly accessible to human senses, thereby enabling us to access and experience them without intermediary equipment, which might also help us gain a more intuitive grasp of their nature \cite{vivoli,sekatski-eyes}. Third, studying macroscopic quantum effects also motivates us to develop a deeper understanding of the notion of macroscopicity itself \cite{frowis}. 

Experimental programs motivated by these goals are currently underway in different physical systems, including atomic ensembles \cite{polzik,kimble,oberthaler,zarkeshian,tiranov,mitchell}, superconducting circuits \cite{friedman,martinis}, molecular interferometers \cite{arndt-c60}, mechanical systems \cite{cleland,lehnert} and light \cite{chekhova,bruno,lvovsky}. There are different levels of macroscopicity that one can strive for in such experiments. One level is to demonstrate quantum phenomena for systems that are macroscopic --- such as ensembles of many atoms, circuits with macroscopic dimensions, or states of light involving many photons, but where the differences between the quantum states that are involved in the superposition or entanglement are still microscopic. Another level is to demonstrate superposition or entanglement that involves individual oblects, but their quantum states are different by macroscopic amounts, and that can therefore be in principle distinguished by detectors that do not have microscopic resolution \cite{sekatski-pra}.

For light, the latter goal has so far been achieved in experiments demonstrating micro-macro entanglement \cite{bruno,lvovsky}. In these experiments, single-photon entanglement was generated by sending a single photon onto a beam splitter. A phase space displacement was then applied to one output mode of the beam splitter to make it macroscopic. In particular, Ref.~\cite{lvovsky}  demonstrated the entanglement of that state, the macroscopic character of its components, and their single-shot distinguishability via detectors with macroscopic resolution. These studies were reminiscent of Schr\"{o}dinger's Gedankenexperiment, in which the living or dead state of a macroscopic cat was entangled with the microscopic state of a radioactive atom.

In the present work, we go one step further and reproduce a situation that is analogous to two cats that are entangled with each other: when one is alive, the other is dead and vice versa, but the state of each individual one is undetermined. We apply macroscopic displacements to \emph{both} channels of a delocalized single photon, thereby obtaining a situation in which two macroscopic and macroscopically distinct states, the displaced single photon (living cat) and the coherent state (dead cat), coexist in an entangled state of two spatially separated optical modes.

\begin{figure}[t]
	\includegraphics[width=\columnwidth]{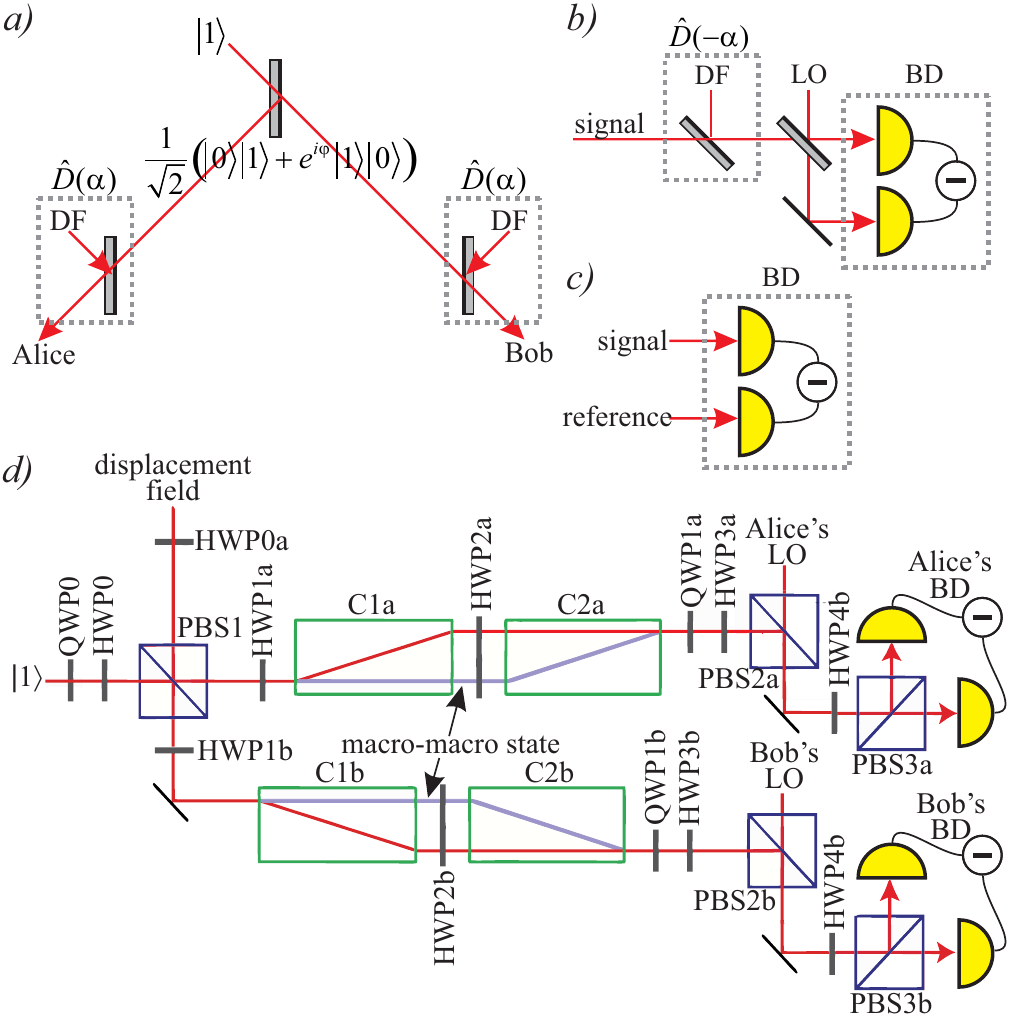}
	\caption{Scheme of the experiment. a) Preparation of the macro-macro entangled state consists in splitting a heralded single photon and applying the phase-space displacement in both modes.  
		b,c) The two ways to characterize the macro-macro entangled state. b) Reverse displacement followed by quadrature measurements on a homodyne detector. c) Photon number measurement using a balanced detector with a reference beam.  d) Full setup. Half-waveplates are denoted by HWP, quarter-waveplates by QWP, polarization beam splitters by PBS, calcite crystals by C, balanced detectors by BD, displacement fields by DF and local oscillator/reference fields by LO. The spatial modes in which the macro-macro entanglement \eqref{eq2} is present are shown by purple lines.}
	\label{f1}
\end{figure}

\paragraph{Concept.} Experimental realization of the macro-macro entanglement begins with the generation of a delocalized photon \cite{Lvovsky1}
\begin{align}
\label{eq1}
\ket{\Psi_0}_{AB}=&\frac{1}{\sqrt{2}}(\ket{0}_{A}\ket{1}_{B}+e^{i\varphi}\ket{1}_{A}\ket{0}_{B})
\end{align}  in Alice's and Bob's modes. We then apply the  phase-space displacement operator $\hat{D}(\alpha)=e^{\alpha\hat a^\dag-\alpha^*\hat{a}}$ (where $\alpha\gg 1$ is the real displacement amplitude) to each mode. This transformation [Fig.~\ref{f1}(a)] converts state (\ref{eq1}) to
\begin{align}\label{eq2}
\ket\Psi_{AB}=&\frac{1}{\sqrt{2}}(\hat{D}(\alpha)\ket{0}_{A}\hat{D}(\alpha)\ket{1}_{B}\\
& +e^{i\varphi}\hat{D}(\alpha)\ket{1}_{A}\hat{D}(\alpha)\ket{0}_{B})\nonumber ,\nonumber
\end{align} 
We implement the displacement by mixing the signal with a strong coherent state on a highly asymmetric beam splitter  \cite{paris,Lvovsky2002}. In both modes, the displacement amplitudes correspond to  $\alpha^2\approx1.1\cdot10^{8}$ photons. 

The states $\hat{D}(\alpha)\ket{0}$ (dead cat) and $\hat{D}(\alpha)\ket{1}$ (living cat), while being similar in mean photon numbers  ($\langle n\rangle_{\hat D(\alpha)\ket 0}=\alpha^2$ and $\langle n\rangle_{\hat D(\alpha)\ket 1}=\alpha^{2}+1$), have significantly different photon number statistics. In particular, their photon number variances  ($\langle \Delta n^2\rangle_{\hat D(\alpha)\ket 0}=\alpha^2$ and $\langle \Delta n^2\rangle_{\hat D(\alpha)\ket 1}=3\alpha^2$) are both macroscopic and different by a factor of 3, so the two states are macroscopically distinguishable. At the same time, the displaced state (\ref{eq2}) exhibits robustness to optical losses at the same level as the state (\ref{eq1}) before displacement  \cite{lvovsky}. This feature makes this state attractive for studying macro-macro entanglement.

We demonstrate the entanglement and macroscopic distinguishability of the two components of state (\ref{eq2}) by making two kinds of measurements. First, both Alice and Bob measure the light pulse energies in their channels, revealing the macroscopic correlations between these energies. Second, we reconstruct the state by means of homodyne tomography after reversing the displacement (hereafter referred to as ``undisplacing") to both modes [Fig.~\ref{f1}(b)] and confirm the entanglement. Displacement is a local operation, therefore the entanglement present in the undisplaced state also proves the entanglement of the macro-macro state. The extra step of undisplacement is necessary because homodyne tomography requires the state being measured to be microscopic. 

To analyze the first type of measurement, we study the  decompositions $\Xi_{0,1}(n)= \bra n\hat{D}(\alpha)\left |0,1\right \rangle$ of the two cat states in the photon number basis \cite{DFS}:
\begin{equation}\label{PsiXi0}
\Xi_{0}(n)=\frac{e^{-\alpha^{2}/2}\alpha^{n}}{\sqrt{n!}};\quad 
\Xi_{1}(n)=\frac{e^{-\alpha^{2}/2}\alpha^{n}}{\sqrt{n!}}\left(\frac{n}{\alpha}-\alpha\right).
\end{equation} 
Their important feature is the behavior of the  ratio
\begin{align}
\left|\frac{\Xi_{1}(n)}{\Xi_{0}(n)}\right|\to	\left\{{0\ {\rm for}\ |n-\alpha^{2}|\ll\alpha}\atop{\infty\ {\rm for}\ |n-\alpha^{2}|\gg\alpha}\right..
\end{align}
Suppose, for example, that Alice observes a photon number that is close to the mean value $\alpha^2$. This indicates that the incoming state in Alice's channel is much more likely to be $\hat D(\alpha)\ket 0$ than $\hat D(\alpha)\ket 1$. Accordingly, Bob's state is $\hat D(\alpha)\ket 1$. On the other hand, if Alice observes her photon number to be far away from the mean, the situation is likely opposite: Alice's cat is alive while Bob's is dead. In both cases, Bob can confirm the state that Alice's measurement prepares in his channel by measuring the (macroscopic) photon number statistics in his mode. 

In the experiment, we measure the pulse energy  by means of a balanced detector. Bob's and Alice's modes are incident upon the sensitive area of one of its photodiodes while the other photodiode is illuminated by a reference pulse of the same mean energy [Fig.~\ref{f1}(c)]. The photocurrents of the two photodiodes are subtracted, yielding the energy of the signal pulse relative to the reference. This technique is necessary in order to eliminate the classical fluctuations of the master laser intensity which would otherwise mask the observation of the quantum effects on photon number variances. The trade-off is an extra unit of shot noise $\left \langle \Delta n^{2}  \right \rangle=\alpha ^{2}$ added to the photon number measurements \cite{lvovsky}.


\begin{figure}
		\includegraphics[width=1\linewidth]{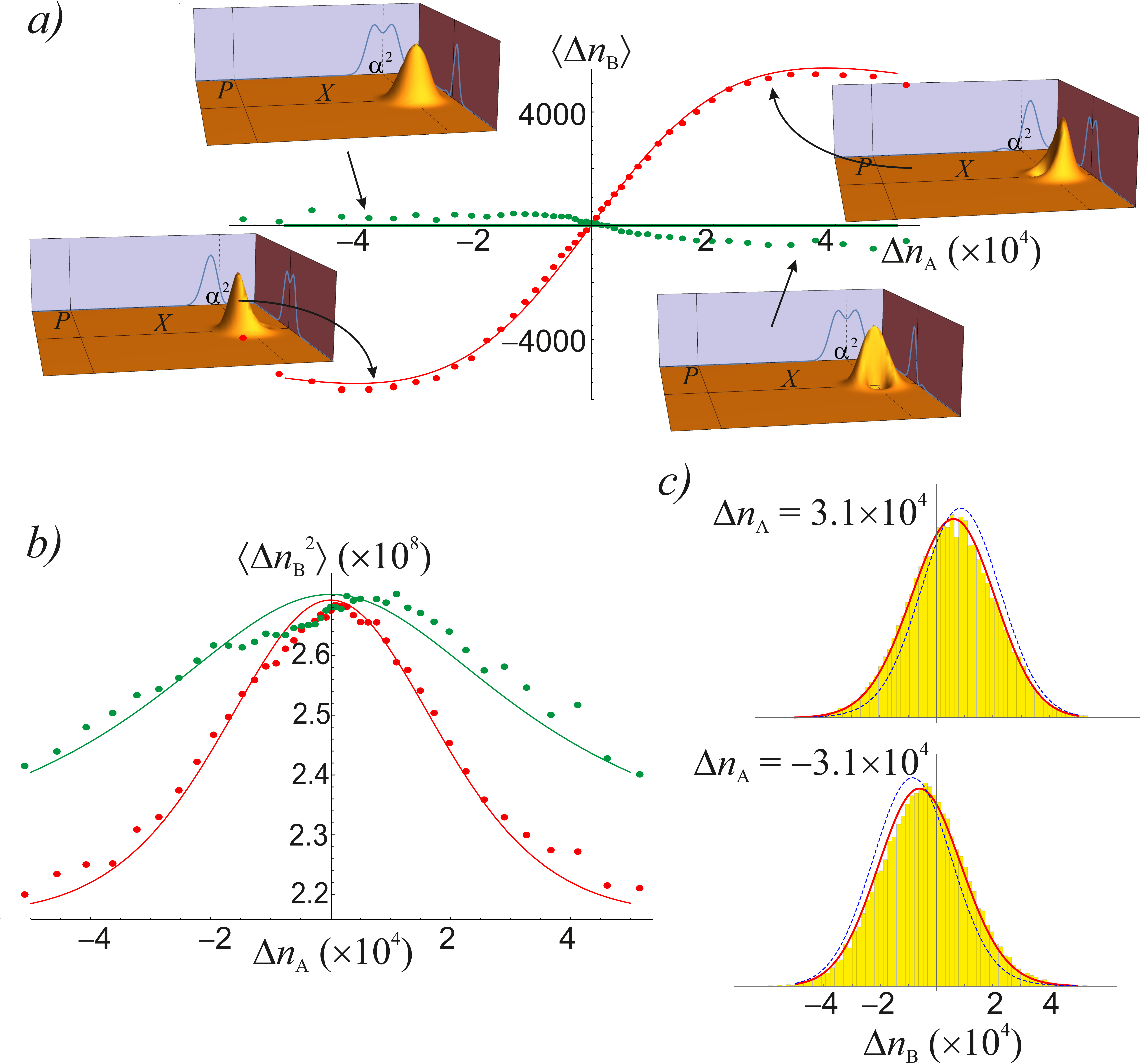}
		\caption{Photon statistics of Bob's mode as a function of energy measurement results in Alice's mode: mean (a), variance (b). The measurements in both channels are relative to the reference pulse. The data corresponding to the relative phase $\varphi=0$ between Alice and Bob is depicted in red color, $\varphi=\pi/2$ in green. Circles show the  experimental data and solid lines are the theoretical curves for the appropriate efficiencies. Insets: Wigner functions and quadrature marginal distributions of Bob's states illustrating why the mean energy of Bob's state depends on Alice's result for $\varphi=0$ but not $\varphi=\pi/2$. c) Histograms of Bob's detected photon numbers conditioned on different results of Alice's measurement, showing a high degree of single-shot distinguishability at a macroscopic level. Red solid line: theoretical prediction; blue dashed line: prediction for 100\%-efficient detectors with the reference beam.}
		\label{f2}
\end{figure}

\begin{figure}[b]
	{\includegraphics[width=0.75\linewidth]{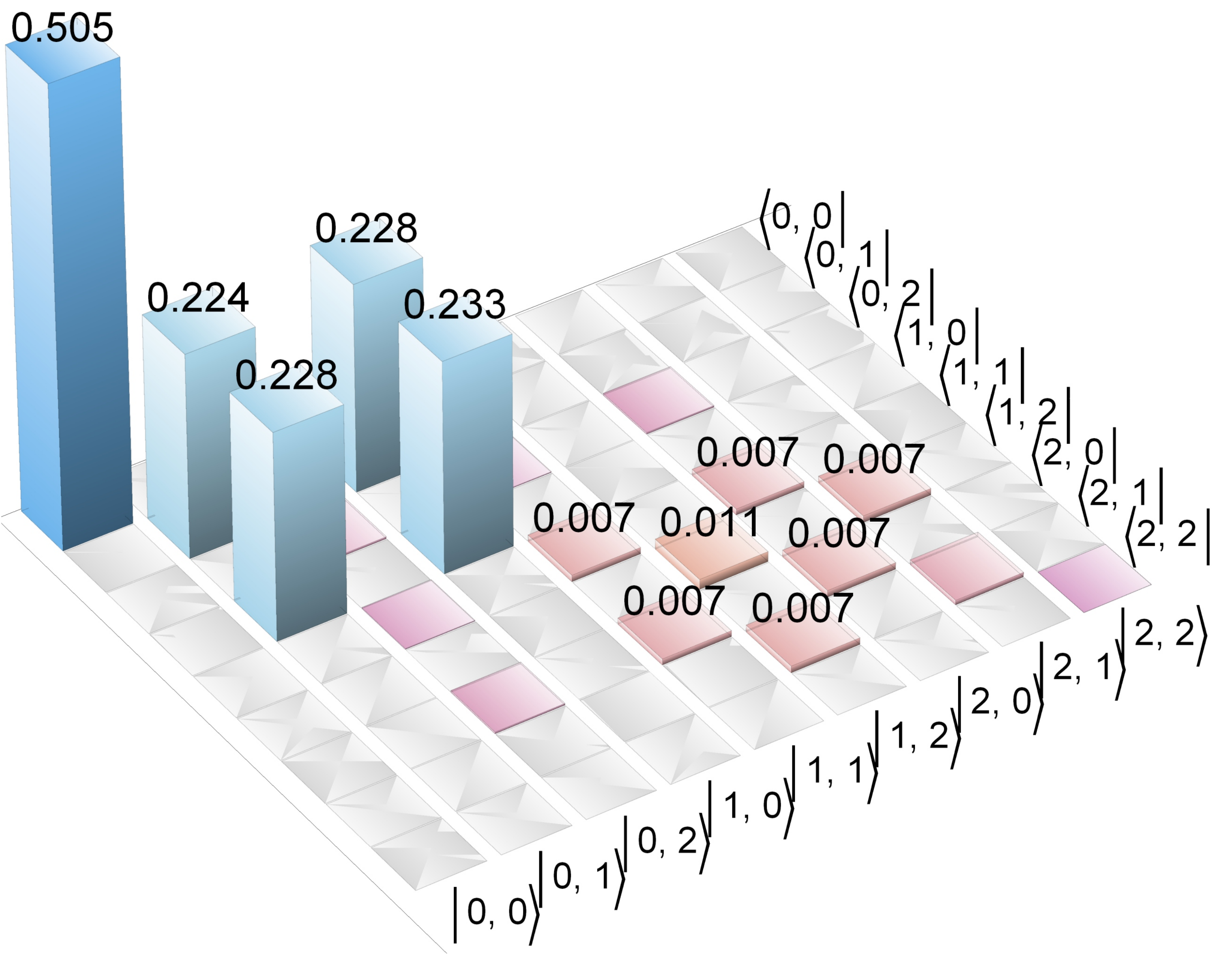} \\}
	\caption{Two-mode density matrix, reconstructed via homodyne tomography after reverse displacements in both channels. The vacuum component appears due to the non-ideal quantum efficiency. Concurrence $C(\rho)=0.32$.}
	\label{f3}
\end{figure}

\paragraph{Experiment.} To prepare the heralded photon, a periodically poled potassium titanyl phosphate (PPKTP) crystal is pumped with 25 mW of frequency-doubled radiation of the master laser --- a Ti:Sapphire Coherent Mira $900$ --- with a wavelength of $780$ nm, a repetition rate of $76$ MHz and a pulse width of $1.5$ ps. As a result of type-II down-conversion, a pair of orthogonally polarized photons (signal and idler) is created. Idler photons are filtered spatially with a single-mode fiber and spectrally with a $~0.2$ nm interference filter after which they are registered by an Excelitas single-photon counting module. Count events occur at a rate of $40-45$ KHz, heralding the preparation of photons in the signal mode \cite{InstantFock}. 

These photons enter the setting shown in Fig.~\ref{f1}(d). First, the photon,  polarized at $45^\circ$, passes through a half-wave plate HWP0 and polarizing beam splitter PBS1, which in combination act as a variable-reflectivity beam splitter. Whenever we wish to change the phase $\varphi$ to $\pi/2$, we insert a quarter wave-plate QWP0 with the optical axis oriented vertically before PBS1.  After PBS1, we have a delocalized single photon (\ref{eq1}). A strong displacement field, also polarized at $45^\circ$, enters the other input port of PBS1 in a matching spatiotemporal mode. In each output after PBS1, the displacement field and the delocalized photon are therefore in orthogonal polarizations of the same spatiotemporal mode. The mean power of the displacement field  just after PBS1 equals $30$ mW in each channel.

The output channels of PBS1 are associated with Alice and Bob. In the further description, we specialize to Alice's setup; Bob's setup is identical. To perform the direct displacement, we apply a half-waveplate (HWP1a) rotated  by $\theta={7}^{\circ}$ with respect to the horizontal axis. Thus, a fraction of the horizontally polarized displacement field with a power of $30\sin^2(2\vartheta)=1.75$ mW is added into the vertical polarization mode of the delocalized photon, thereby performing the displacement by $\alpha=1.05\cdot10^{4}$. The two polarization modes are then spatially separated in a 40-mm birefringent calcite crystal (C1a). A half-waveplate HWP2a rotates both polarizations by $90^\circ$ after which they are spatially recombined by another calcite crystal (C2a) identical to the first one. Care is taken to ensure that the optical paths in Alice's and Bob's channel from PBS1 to the gap between the two calcite crystals are equal, which means that both channels are macroscopic at the same time.   


The amplitude and the phase of the reverse displacement are determined by the half-wave plate HWP3a and the quarter-wave plate QWP1a. By adjusting them, we either completely undisplace the signal by means of destructive interference between the signal and the displacement field on PBS2a or keep the amplitude of the second displacement equal to zero. For high-quality reverse displacement that would enable homodyne detection of the signal, high phase stability of the interferometer formed within the two calcite crystals is essential. This was achieved by high mechanical stability of the optical mounts and by keeping the size of the interferometer to a minimum.

The photon number and quadrature detection [i.e.~the detection methods of Figs.~\ref{f1}(b,c)] are implemented using the same apparatus formed by PBS3a and HWP4a. For homodyne detection, the half-wave plate HWP4a is set to rotate the polarizations by $45^\circ$ so PBS3b acts as a symmetric beam splitter that mixes the LO and signal and sends them onto the two photodiodes of the balanced detector (BD) \cite{Masalov2017}. In the case of photon number statistics measurement, the local oscillator plays the role of the reference beam. In this case, HWP4a is set to $0^\circ$, so the signal and the reference beam go onto different diodes of the BD without mixing.

For the homodyne tomography experiment, we must know the relative phases between the signals and local oscillators at any moment in time. This phase is measured  using the interference signal between the local oscillator and the displacement field transmitted through PBS2.  The phase in Bob's mode is actively stabilized by means of a piezoelectric transducer. The phase in Alice's mode was varied from 0 to $2\pi$ and recorded.

The total quantum efficiency is affected by a non-ideal mode matching between the local oscillators and the signals ($81\%$), losses in all optical elements including the uncoated calcite crystals ($77\%$) and the homodyne detector's quantum efficiency ($86\%$). Additional $\sim5$\% losses in the tomographic measurement arise due to imperfect modematching between the signal and undisplacement field. For the photon number measurements the efficiency is decreased by a similar fraction, but for a different reason:  because the power of both beams incident on the homodyne detector is lower than that typically used in homodyne measurements, the detector's electronic noise plays a more significant role \cite{ElectronicNoise}. The total efficiency in both regimes is about $\eta=49\%$.  

\paragraph{Results and discussions.} The correlations between photon number statistics, calculated from 5,000,000 data samples acquired for each setting, are presented in Fig.~\ref{f2}(a,b). We see that the photon number variances for Bob's mode behave consistently to the discussion above: they are higher when Alice's measured photon number is close to its mean value. The influence of the above-mentioned imperfections and the usage of the reference beam decreased the ratio of the variances from the theoretical expectation of $3$ to a measured value of $(4+\eta)/(4-\eta)=1.33$ \cite{supp}. 

Photon number correlations depend on the relative phase $\varphi$ between the two channels. This is particularly evident for Bob's mean photon number $\langle n_B\rangle$ plotted as a function of  Alice's measurement results. A strong dependence is present for $\varphi=0$, while there is almost no dependence for $\varphi=\pi/2$ [Fig.~\ref{f2}(a)]. This can be understood  by writing the conditionally prepared state in Bob's mode \cite{Lvovsky3}:
\begin{align}\label{eq4}
\langle n_A |\Psi\rangle_{AB} =&\Xi_{1}(n_{A})\hat{D}(\alpha)\ket{0}_{B}+e^{i\varphi}\Xi_{0}(n_{A})\hat{D}(\alpha)\ket{1}_{B}
\end{align} for Alice's photon number measurement with the result $n_A$. This state is a displaced superposition of the vacuum and single-photon states. We recall that the energy of a harmonic oscillator equals 
\begin{equation}\label{energy}\hat H=\frac12\hbar\omega (\hat X^2+\hat P^2)=\frac12\hbar\omega (\hat X_0^2+2X_0\Delta \hat X+\Delta \hat X^2+\hat \Delta P^2),\nonumber
\end{equation}
where we defined $\Delta \hat X=\hat X-X_0$, $\Delta \hat P = \hat P$ with $X_0=\alpha\sqrt 2\gg 1$ being the macroscopic phase-space displacement along the position axis. In the state (\ref{eq4}), the observables  $\Delta \hat X$ and $\Delta \hat P$ are on a scale of $1$, hence the quantum fluctuations of the energy are primarily determined by the macroscopic term $2X_0\Delta \hat X$, which, in turn, is proportional to the quantum fluctuations of the position observable. The fluctuations of the momentum, in contrast, do not affect the observable energy significantly. 

The inset in Fig.~\ref{f2}(a) shows the Wigner functions and marginal distributions of the state (\ref{eq4}). We see that,   for  $\varphi=0$, the position marginal distribution is asymmetric around $X_0$ and its center of mass $\langle\Delta X\rangle$ shifts with changing $n_A$. In contrast,  for $\varphi=\pi/2$, this distribution is always symmetric with  $\langle\Delta X=0\rangle$, resulting in a constant $\langle n_B\rangle$.

Fig.~\ref{f2}(c) shows the histogram of Bob's detected photon numbers for Alice's detected photon numbers being above (top) and  below (bottom) its mean value $\alpha^2$ by  $\delta n_A\sim\pm 3.1\cdot 10^4$. These measurements project Bob's mode onto the states approximating $\hat{D}(\alpha)(\ket{0}\pm\ket{1})$, respectively. As evidenced by the figure, the photon statistics of these states are not only macroscopic, but also substantially different. They can be distinguished through a single measurement with an error probability of 36\%. 

The significance of this observation is emphasized if we rewrite the state \eqref{eq2} for $\varphi =0 $ as 
\begin{align}\label{eq2r}\nonumber
\ket\Psi_{AB}=&\frac{1}{2\sqrt{2}}\big[\hat{D}(\alpha)(\ket{0}+\ket{1})_{A}\hat{D}(\alpha)(\ket{0}+\ket{1})_{B}\\
& +\hat{D}(\alpha)(\ket{0}-\ket{1})_{A}\hat{D}(\alpha)(\ket{0}-\ket{1})_{B}\big].
\end{align} 
The macroscopic, single-shot distinguishability of the two terms of this state is a signature of a macroscopic quantum superposition.



The effect of the phase on the measurable physical properties of the two-mode state, observed in Fig.~\ref{f2}(a,b), constitutes indirect evidence that the quantum coherence between its two terms is present in spite of its macroscopic nature. To obtain more direct evidence, we perform homodyne tomography of the two-mode state after applying reverse phase-space displacement to both modes  [Fig.~\ref{f1}(c,d)]. We record a total of 200,000 quadrature samples in each mode. The phase difference between the two local oscillators is varied, allowing us to get the complete data set for tomography \cite{Lvovsky1}. The two-mode density matrix [Fig.~\ref{f3}] is reconstructed by means of the maximum-likelihood algorithm \cite{Lvovsky2} and displays a statistical mixture of state (\ref{eq2}) and vacuum, arising, again, due to various losses as well as the quadrature noise associated with the combination of direct and reverse displacements. The entanglement is evidenced by the presence of non-diagonal matrix elements \cite{Morin} and can be  quantified in terms of concurrence $C[\hat{\rho}]=2\cdot(\left|\rho_{01}\right|-\sqrt{\rho_{00}\cdot\rho_{11}})$ \cite{concurrence}, which is equal to $0.32$.  As discussed below, the local nature of undisplacement guarantees that the macroscopic state prior to the undisplacement possesses at least the same amount of entanglement. 

\paragraph{Summary.} We have demonstrated the creation of entanglement between macroscopically distinct states of light. It was shown in Ref.~\cite{vivoli} that the present type of quantum state is suitable for demonstrating entanglement using human eyes of detectors \cite{sekatski-eyes}, opening up one fascinating avenue for future experiments. The present approach could also be adapted to the entanglement between atomic ensembles \cite{kimble}. Another interesting direction would be to map the present macro-macro entanglement of light onto mechanical systems, similarly to what was proposed for micro-macro entanglement in Ref.~\cite{Ghobadi2014}, which would open up the possibility of testing wave function collapse models motivated by the quantum measurement problem and by quantum gravity considerations.


\begin{acknowledgements}
The work was supported by the RFBR (Grant No. 18-37-20033). AL and CS's research has been supported by NSERC. AL is a CIFAR Fellow.
\end{acknowledgements}

\appendix
\section{Supplementary information:\\Theoretical predictions for the data in Fig.~2}
 The two-mode  delocalized photon state which is displaced and subjected to losses in each channel can be written as $\hat\rho=\eta\ketbra\Psi\Psi+(1-\eta)\ketbra{0,0}{0,0}$,
where $\eta$  is the efficiency and $\ket\Psi$ is given by Eq.~\eqref{eq2}. The probability of observing $n_A$ photons by Alice and $n_B$ photons by Bob is then
\begin{align}\label{prmn}
{\rm pr}(n_A,n_B)&=\bra{n_A,n_B}\hat\rho_{AB}\ket{n_A,n_B}\\&=\nonumber
\frac\eta2\left|\Xi_0(n_A)\Xi_1(n_B)+e^{i\varphi}\Xi_1(n_A)\Xi_0(n_B)\right|^2\\
&+(1-\eta)\Xi_0^2(n_A)\Xi_0^2(n_B).\nonumber
\end{align}
We approximate the Poissonian distribution by Gaussian for $\alpha\gg1$: $$\frac{e^{-\alpha^2}\alpha^{2n}}{n!}\approx\frac{e^{-\delta n^2/2\alpha^2}}{\sqrt{2\pi}\alpha},$$  
where $\delta n = n-\alpha^2$. Using this approximation and Eq.~\eqref{PsiXi0}, we rewrite Eq.~\eqref{prmn} as follows:
\begin{align}\label{eq61}
&{\rm pr}(n_A,n_B)\\&\nonumber =  \frac{e^{-\frac{\delta n_A^2+\delta n_B^2}{2\alpha^{2}}}}{2\pi\alpha^{4}}  \bigg[\frac{\eta}{2}\big( \delta n_A^2 + \delta n_B^2 + 
2\cos\varphi \delta n_A\delta n_B \big)\\&+(1-\eta)\alpha^2\bigg].\nonumber  
\end{align}
We now recall that our photon number measurements are with respect to the reference pulses. The latter are in coherent states with the photon number distribution ${\rm pr}_{\rm ref}(n_{\rm ref})=\frac1{\sqrt{2\pi}\alpha}e^{-\delta n_{\rm ref}^2/2\alpha^2}$. The probability distribution for the number of electrons in the subtraction photocurrent pulse is obtained by convolving ${\rm pr}(n_A,n_B)$ with the reference distribution both in Alice's and Bob's channels, yielding
\begin{align}\label{eq61conv}
&{\rm pr}'(n_A,n_B)\\&\nonumber =  \frac{e^{-\frac{n_A^2+ n_B^2}{4\alpha^{2}}}}{32\pi\alpha^{4}}  \bigg[\eta\big( n_A^2 +  n_B^2 + 
2\cos\varphi  n_A n_B \big)\\&+4(2-\eta)\alpha^2\bigg].\nonumber  
\end{align}
The theoretical curves in Fig.~2 are the mean and variance of the above, interpreted as Bob's single-channel probability density conditioned on the detection of a specific photon number by Alice. Explicitly:
\begin{align}
\langle{n_B}\rangle&=\frac{4 \alpha^2n_A \eta \cos\varphi 
}{\eta(n_A^2 -2 \alpha^2 ) +8\alpha^2};\\
\langle{\Delta n_B}^2\rangle&=\frac{2 \alpha^2 \left(2 \alpha^2 (\eta
	+4)+n_A^2 \eta \right)}{2 \alpha^2 (4-\eta)+n_A^2 \eta
}-\frac{16 \alpha^4  n_A^2 \eta
	^2\cos^2\varphi}{\left(2 \alpha^2  (4-\eta)+ n_A^2 \eta\right)^2 }.
\end{align}
The peak value of the variance is given by 
$$\frac{\langle{\Delta n_B}^2\rangle|_{n_A=0}}{\langle{\Delta n_B}^2\rangle|_{n_A\to\infty}}=\frac{4+\eta}{4-\eta},$$
which equals to $1.28$ for the experimental efficiency of $\eta=0.49$. The actual ratio observed in the experiment is $1.33$.

\end{document}